# Sardinia Radio Telescope: General Description, Technical Commissioning and First Light

P. Bolli[*], A. Orlati[†], L. Stringhetti[‡,¶], A. Orfei[†], S. Righini[†], R. Ambrosini[†],
M. Bartolini[†], C. Bortolotti[†], F. Buffa[§], M. Buttu[§], A. Cattani[†], N. D'Amico[§],
G. Deiana[§], A. Fara[§], F. Fiocchi[†], F. Gaudiomonte[§], A. Maccaferri[†], S. Mariotti[†],
P. Marongiu[§], A. Melis[§], C. Migoni[§], M. Morsiani[†], M. Nanni[†], F. Nasyr[§], A. Pellizzoni[§],
T. Pisanu[§], M. Poloni[†], S. Poppi[§], I. Porceddu[§], I. Prandoni[†], J. Roda[†], M. Roma[†],
A. Scalambra[†], G. Serra[§], A. Trois[§], G. Valente[§], G. P. Vargiu[§] and G. Zacchiroli[†]

[*]INAF – Osservatorio Astrofisico di Arcetri, Florence, Italy

[†]INAF – Istituto di Radio Astronomia di Bologna, Bologna, Italy

[‡]INAF – Istituto di Astrofisica Spaziale e Fisica Cosmica di Milano, Milan, Italy

[§]INAF – Osservatorio Astronomico di Cagliari, Cagliari, Italy

[¶]luca@iast.milano.inaf.it

In the period 2012 June–2013 October, the Sardinia Radio Telescope (SRT) went through the technical commissioning phase. The characterization involved three first-light receivers, ranging in frequency between 300 MHz and 26 GHz, connected to a Total Power back-end. It also tested and employed the telescope active surface installed in the main reflector of the antenna. The instrument status and performance proved to be in good agreement with the expectations in terms of surface panels alignment (at present 300 μm rms to be improved with microwave holography), gain (∼0.6 K/Jy in the given frequency range), pointing accuracy (5 arcsec at 22 GHz) and overall single-dish operational capabilities. Unresolved issues include the commissioning of the receiver centered at 350 MHz, which was compromised by several radio frequency interferences, and a lower-than-expected aperture efficiency for the 22-GHz receiver when pointing at low elevations. Nevertheless, the SRT, at present completing its Astronomical Validation phase, is positively approaching its opening to the scientific community.

## 1. Introduction

The Sardinia Radio Telescope (SRT) (Lat. 39°29′34″N – Long. 9°14′42″E; 600 m above the sea level) is a new Italian facility for radio astronomy whose commissioning was completed at the end of 2013. Its formal inauguration took place on 2013 September 30. The antenna is a fully steerable, wheel-and-track dish, 64 m in diameter, located 35 km north of Cagliari, on the island of Sardinia. It completes a set of three antennas devoted to radio astronomical science in Italy (Fig. 1), all managed by the Italian Institute for Astrophysics (INAF).

The SRT is a general-purpose radio telescope aimed at operating with high aperture efficiency. Once all the planned devices are installed, it will observe in the frequency range from 300 MHz to 100 GHz and beyond (from 1 m to 3 mm in wavelength). The antenna gain is expected to vary from 0.50 to 0.70 K/Jy for the frequency range 0.3–50 GHz and to be around 0.34 K/Jy in the 3 mm band (70–115 GHz).

A key feature of the SRT is its active surface, in total composed by 1116 electromechanical actuators, able to correct the deformations induced by gravity on the primary surface (or "mirror"). Effort is underway to employ this facility to also correct

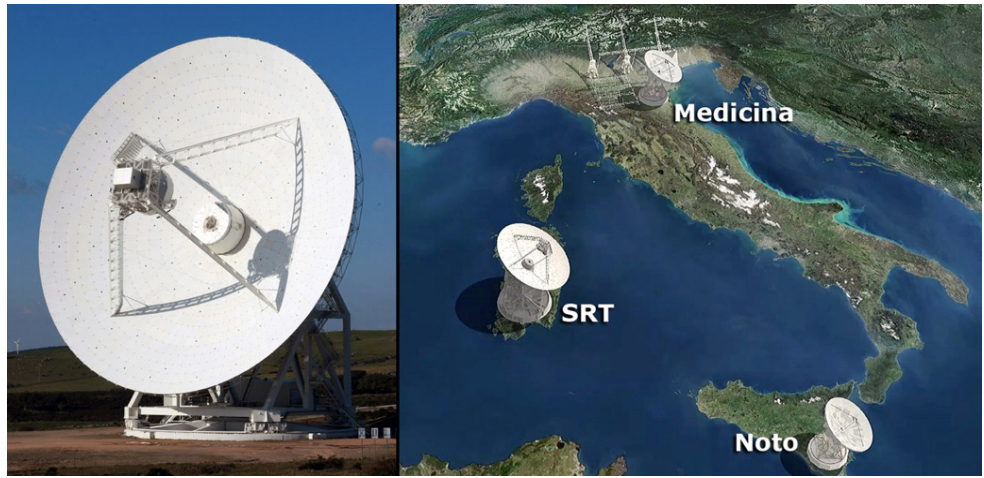

Fig. 1. The SRT and the location of the three antenna sites.

for non-systematic errors, such as temperature/wind-related effects.

The SRT is capable of hosting many microwave receivers, located in four different antenna focal positions — primary, secondary and two beam-waveguide foci — able to cover almost continually its frequency range. The SRT will operate in single-dish (continuum, full Stokes and spectroscopy), Very Long Baseline Interferometry (VLBI) and Space Science (Ambrosini, 2011a) modes.

Thanks to its large aperture and versatility (multi-frequency agility and wide frequency coverage), the SRT is expected to have a major impact in a wide range of scientific areas for many years to come. A full description of the potential SRT science applications is beyond the scopes of the present paper, as it is provided in a separate paper dedicated to the Astronomical Validation activities (Prandoni *et al.*, in preparation). Here we illustrate some of the main areas where we think the SRT can play a major role in the next future.

Operations in the framework of international VLBI and Pulsar Timing networks are of top priority for the SRT. SRT is going to be one of the most sensitive European VLBI Network (EVN) stations, together with Effelsberg and Jodrell Bank. Its large aperture is also of extreme importance for Space VLBI observations with RadioAstron. Thanks to its active surface, the SRT will also represent a sensitive element of the mm-VLBI network operating at 7- and 3-mm bands, where a substantial improvement in collecting area and in the coverage of the sky Fourier transform plane is of vital importance for increasing the number of targets accessible to the array and the quality of the images. Once the fiber optic connection to the site is completed, the SRT will also participate in real-time VLBI observations (e-VLBI). The availability of three antennas (Fig. 1) will moreover allow the constitution of a small Italian VLBI network, exploiting a software correlator (DiFX, already operating). The SRT will be also included in the geodetic VLBI network.

The SRT is one of the five telescopes of the European Pulsar Timing Array (EPTA), which, together with the North American Nanohertz Observatory for Gravitational Waves (NANOGrav), and the Parkes Pulsar Timing Array (PPTA) share the goal to detect gravitational waves. Being the southernmost telescope of the EPTA collaboration, the SRT will allow a better coverage of pulsars with declinations below $-20°$, hence a better overlap with the PPTA. Thanks to its dual-band L/P receiver, SRT will be of great importance in measuring accurate dispersion measure variations, crucial to obtain ultra-precision pulsar timing, and search for signatures of space-time perturbations in the pulsar timing residuals. The SRT is also part of the Large European Array for Pulsars (LEAP), a project which consists in using the EPTA telescopes in tied-array mode, obtaining the equivalent of a fully steerable 200-m dish.

The SRT is expected to have a major impact also for single-dish observations. In particular, we aim at exploiting its capability to operate with high efficiency at high radio frequency. Equipped with multi-feed receivers the SRT can play a major role in conducting wide-area surveys of the sky in a frequency range (20–90 GHz), which is poorly explored, yet very interesting.

For instance the first-light K-band 7-beams receiver will be exploited to obtain extensive mapping of the NH3 in Galactic star-forming regions, in close synergy with existing IR/sub-mm continuum surveys of the Galactic Plane. The ammonia molecule, through its (1,1) and (2,2) transitions, is a good tracer of dense cores of molecular gas and is considered an excellent thermometer. Another interesting application of the K-band multi-feed receiver is a H2O line survey of nearby galaxies. This will increase the number of detected water masers and derive distances, dynamical models and total masses of (luminous + dark) matter of the galaxies in the Local Group.

Wideband multi-feed receivers operating at higher frequency (40–90 GHz) — currently under development — will allow us to get access to unique molecular line transitions in our own Galaxy, like for example those associated with deuterated molecules (e.g. DCO + (1−0), N2D + (1−0)), crucial to constrain the kinematic and chemical properties of pre-stellar cores, as well as to uncover the cool molecular content of the Universe in a crucial cosmic interval (redshift ∼0.3–2), through the mapping of redshifted CO low-J transitions. Wideband radio-continuum (and spectro-polarimetry) mapping of extended (low-surface brightness) Galactic and extra-galactic sources (e.g. SNRs, radio galaxies, nearby spirals), on the other hand, will permit resolved studies aimed at a better understanding of the physics of accretion and star formation processes. In addition high-frequency receivers will allow us to conduct (high spatial resolution) follow-up observations and monitoring experiments of AGNs, GRBs and other transient events in connection with high-energy experiments (FERMI, MAGIC, CTA), a research field where the Italian astronomical community is very active.

Due to an agreement between INAF and ASI on the use of the instrument for space applications, SRT will also be involved in planetary radar astronomy and space missions (Tofani *et al.*, 2008; Grueff *et al.*, 2004).

This paper presents an overview of the SRT system in the light of the results obtained during the commissioning phase (Ambrosini *et al.*, 2013). Section 2 describes the overall SRT system, including technical specifications of the main antenna modules. Section 3 shows the results of the tests performed in the commissioning phase and the timeline for the main milestones. In Sec. 4 some conclusions are drawn on the overall results of the commissioning.

For the convenience of readers, Table 1 provides a glossary of many acronyms used throughout the paper.

Table 1. Glossary of the terms used.

| Acronym | Description |
|---|---|
| ACS | Alma Common Software |
| ACU | Antenna Control Unit |
| ASI | Agenzia Spaziale Italiana |
| AV | Astronomical Validation |
| BWG | Beamwaveguide |
| EPTA | European Pulsar Timing Array |
| DBBC | Digital Baseband Converter |
| DFB | Digital Filter Bank |
| DiFX | Distributed Fourier spectrum Cross-multiplied |
| DISCOS | Development of the Italian Single-dish Control System |
| DSP | Digital Signal Processing |
| EER | Elevation Equipment Room |
| e-VLBI | Real-time VLBI |
| EVN | European VLBI Network |
| FEM | Finite Element Method |
| FOV | Field of View |
| HPBW | Half-Power Beam Width |
| IF | Intermediate Frequency band |
| INAF | Istituto Nazionale di Astrofisica |
| IRA | Istituto di Radioastronomia |
| IRIG-B | Inter-Range Instrumentation Group, time code format B |
| LCP | Left Circular Polarization |
| LEAP | Large European Array for Pulsar |
| OAA | Osservatorio Astrofisico di Arcetri |
| OAC | Osservatorio Astronomico di Cagliari |
| OTF | On-The-Fly |
| PFP | Primary Focus Positioner |
| PSD | Position Sensing Device |
| RCP | Right Circular Polarization |
| RFI | Radiofrequency Interference |
| ROACH | Reconfigurable Open Architecture and Computing Hardware |
| SDI | Single-Dish Imager |
| SRT | Sardinia Radio Telescope |
| T&F | Time and Frequency |
| VLBI | Very Long Baseline Interferometry |
| WVR | Water Vapor Radiometer |
| XARCOS | Arcetri Cross-Correlator Spectrometer |

## 2. The Sardinia Telescope System

### 2.1. *Antenna technical characteristics*

The antenna design, whose schematic view is shown in Fig. 2, is based on a wheel-and-track configuration. The main reflector (M1) consists of a back-structure that supports — through actuators — the mirror surface, itself composed of rings of reflecting panels. A quadrupod, connected to the back-structure, supports the sub-reflector (M2) and the primary focus positioner and instrumentation. The main and secondary mirrors and the quadrupod lie on the alidade, which is a welded steel structure standing on a large concrete tower that forms the bulk of the antenna foundation. Three large rooms were built behind the primary mirror to contain the secondary focus receivers, the beam-waveguide mirrors and several electronic instrumentation and cable distributions, respectively. A fourth room is located in the lower part of the alidade, where the power drivers for the motors, the antenna control unit and the cryogenic compressors are installed. The alidade also supports the elevation wheel, which is a conical truss anchored to the reflector back-structure through a massive pyramidal structure.

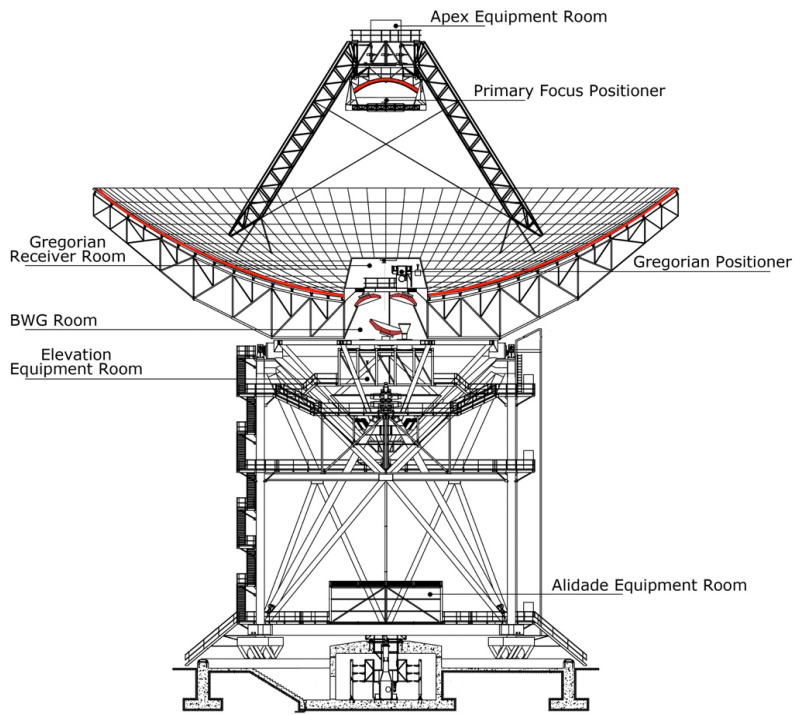

Fig. 2. SRT: mechanical structure. The radio-frequency reflecting surfaces are highlighted in red.

The radio telescope is supported by a reinforced-concrete foundation excavated into rock; an outer ring beam sustains the azimuth track, while a central building accommodates the azimuth pintle bearing support, the azimuth cable wrap and the encoder system. The wheel-and-track structure consists of 16 wheels lying on the rail, which is a continuous welded ring 40 m in diameter. This is connected to the foundation by 260 pairs of anchor bolts and by a fiber-reinforced grout. The overall structure of the antenna weighs approximately 3300 tons.

The antenna is steerable around the azimuth and elevation axes. The rotation around the two axes is managed by a servo control system, consisting in 29-bit absolute encoders (peak-to-peak position error 0.8 arcsec), 12 brushless asynchronous motors (eight in azimuth and four in elevation) and an ACU available on a BECKHOFF hardware platform employing an IRIG-B generator. A proper torque bias is applied to the motors to overcome the gearbox backlash and to improve the antenna pointing accuracy.

The primary reflector surface consists of 1008 individual aluminum panels divided into 14 rows of identical panel types. Each panel has an area ranging from 2.4 to $5.3\,\mathrm{m}^2$. It is built using aluminum sheets glued, by means of a layer of epoxy resin, to both longitudinal and transversal Z-shaped aluminum stiffeners. The basic back-structure is composed of 96 radial trusses and 14 circumferential trussed hoops supported by a large center hub ring. The sub-reflector surface consists of 49 individual aluminum panels with an average area of about $1\,\mathrm{m}^2$, whereas its back-structure is formed by 12 radial trusses and three circumferential trussed hoops supported by a center hub ring. Three of these trusses are directly connected to a triangular steel frame, which has the function of a transitional structure to the six sub-reflector actuators. These actuators define the sub-reflector position, and provide for sub-reflector motion with five degrees of freedom.

Further mirrors beneath the Gregorian focus, arranged in a Beam WaveGuide configuration, allow the addition of four more focal points with magnified and de-magnified $F/D$ ratio in the intermediate frequency bands (Fig. 3).

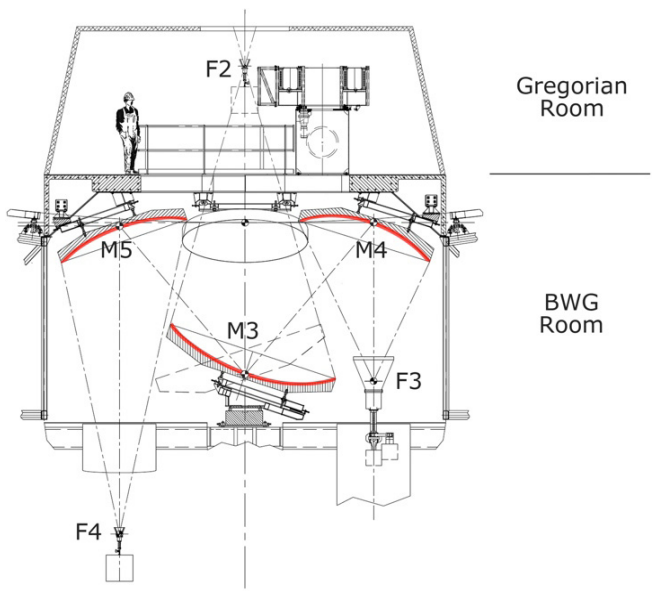

Fig. 3. Technical view of the Gregorian and BWG rooms containing the Gregorian turret and the BWG mirrors, respectively.

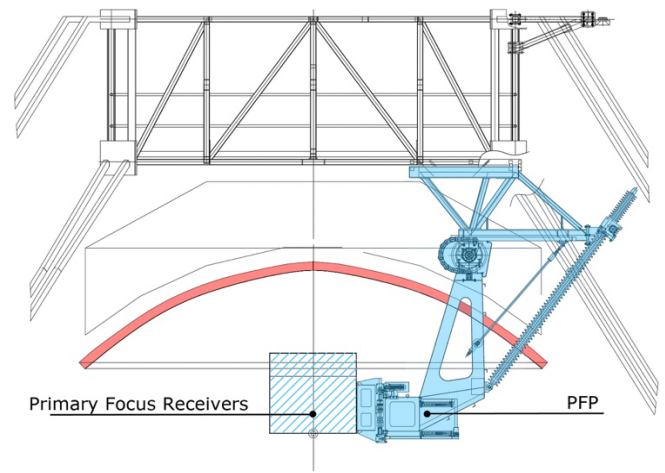

Fig. 4. Technical view of the primary focus area. The primary focus positioner, containing the primary focus receivers, and the reflecting surface of the sub-reflector are highlighted in light blue and red, respectively.

Two of the four designed BWG layouts have already been constructed. They use three mirrors: M3 as a shared mirror, M4 for the first layout and M5 for the second one. The re-imaging optics for BWG *layout I* were designed for maximum focal ratio reduction (right side of Fig. 3); whereas BWG *layout II* was designed so that the output focal point F4 lies beneath the elevation axis of the antenna (left side of Fig. 3). By an opportune rotation of M3, which takes 1.5 min, the desired BWG layout is selected. All the mirrors are portions of ellipsoids and present quite large apertures, as shown in Table 2. The two missing BWG layouts will be added in a later stage to the present configuration using two more mirrors and an appropriate rotation of M3; they will be dedicated to Space Science applications.

A rotating turret (Gregorian positioner assembly) is mounted eccentrically on the focal plane of the antenna dish, and can house eight separate cryogenic receiving systems and the associated feed horns for operating up to 115 GHz. A drive system can rotate the turret (within 2 min for a complete rotation of 340°) so that any of the feed horns can be positioned on the focal plane. The servo control system consists of two brushless servo motors with drivers and a position-control computer.

A mechanism selecting among different receivers exists also in the primary focus. Several receiver assemblies, the exact number depending on their dimensions, can be allocated at the secondary mirror back-structure side. An arm controlled by two servo motors can, in less than 4 min, place the required receiver box in the primary focus position (Fig. 4), which can accommodate low-frequency receivers whose dimensions are not mechanically compatible with the Gregorian positioner assembly.

The remote and automatic control of all these movements makes the antenna available in frequency agility, switching among all the observing bands in a fast and unmanned way.

Suitable atmospheric conditions (see Table 2 for the environmental specifications) together with the active surface will extend the use of the instrument up to 100 GHz. This capability also requires an advanced metrology system for accurate antenna pointing (see Sec. 2.3). The availability of a Water Vapor Radiometer (WVR) and a Weather Station, already in operation, will also allow the dynamic scheduling of the observations, further improving the telescope productivity.

A complex helium 5.5 (99.9995% pure gas) plant consisting of seven pairs of supply-and-return lines assures that the microwave receivers, distributed in the various positions, can operate at cryogenic temperatures. The system is dimensioned so that each line can serve up to three microwave receivers. The total length of the lines reaches 1.5 km, and they are composed of both rigid and flexible tubes.

Table 2. Specifications of the SRT.

| Characteristics | Details | Value |
| --- | --- | --- |
| Range of motion | Elevation | 5–90° |
|  | Azimuth | 180° ± 270° |
| Slew rate | Elevation | 0.5°/s |
|  | Azimuth | 0.85°/s |
| Mirrors size | M1 – Primary mirror | 64 m (axially symmetrical) |
|  | M2 – Sub-reflector | 7.9 m (axially symmetrical) |
|  | M3 – BWG shared between *layouts I* and *II* | 3.921 × 3.702 m |
|  | M4 – BWG *layout I* | 3.103 × 2.929 m |
|  | M5 – BWG *layout II* | 2.994 × 2.823 m |
| Number of individual panels | M1 – Primary mirror | 1008 (14 circular rows) |
|  | M2 – Sub-reflector | 49 (3 circular rows) |
| Expected surface accuracy (El = 45°) in precision environment conditions | M1 panel | ≤ 65 micron rms |
|  | M1 alignment (photogrammetry; holography) | 290; 150 micron rms |
|  | M2 panel | ≤ 50 micron rms |
|  | Back-up structure (with active surface) | 0 |
|  | Actuator accuracy and linearity | 15 micron rms |
|  | Other (meas. errors; thermal and wind effect) | 49 micron rms |
|  | Total without holography | 305 micron rms |
|  | Total with holography | 178 micron rms |
| Pointing accuracy | Normal conditions without metrology systems | 13 arcsec |
|  | Precision conditions with metrology systems | 2 arcsec |
| F/D | F1 – Primary focus | 0.33 |
|  | F2 – Gregorian focus | 2.34 |
|  | F3 – BWG *layout I* focus | 1.37 |
|  | F4 – BWG *layout II* focus | 2.81 |
| Frequency range | Primary focus | 0.3–20 GHz |
|  | Gregorian focus | 7.5–115 GHz |
|  | BWG foci | 1.0–35 GHz |
| Precision environment conditions | Wind | < 15 km/h |
|  | Solar | Absent |
|  | Precipitation | Absent |
|  | Temperature | −10°C to 30°C |
|  | Temperature drift | < 3°C/h |
|  | Humidity | < 85% |
| Normal environment conditions | Wind | < 40 km/h |
|  | Solar | Clear sky |
|  | Precipitation | Absent |
|  | Temperature | −10°C to 40°C |
|  | Temperature drift | < 10°C/h |
|  | Humidity | < 90% |

For the radio frequency links (Fig. 5), more than 3 km of coaxial cables connect the focal positions to a central point located in the elevation Equipment Room. The highest frequency useable with the co-axial cables is 4 GHz, with a maximum attenuation of 15 dB (at 2 GHz) for the longest path (100 m) and with a matching coefficient better than −20 dB over the whole frequency band. An optical fiber cabling, based on both single-mode and multi-mode fibers, was deployed within the radio telescope for digital and analog data transmission. The optical fibers are responsible for the connection of the radio telescope to the external infrastructures. Some customized optical links allow the transport of the astronomical signal, in the IF frequency band (0.1–2.1 GHz), from the EER to the remote control and data processing room. This solution, thanks to a link gain of 0 dB associated with a dynamic range better than the dynamic range of the receiver itself, ensures that the overall specifications of any receiver are maintained.

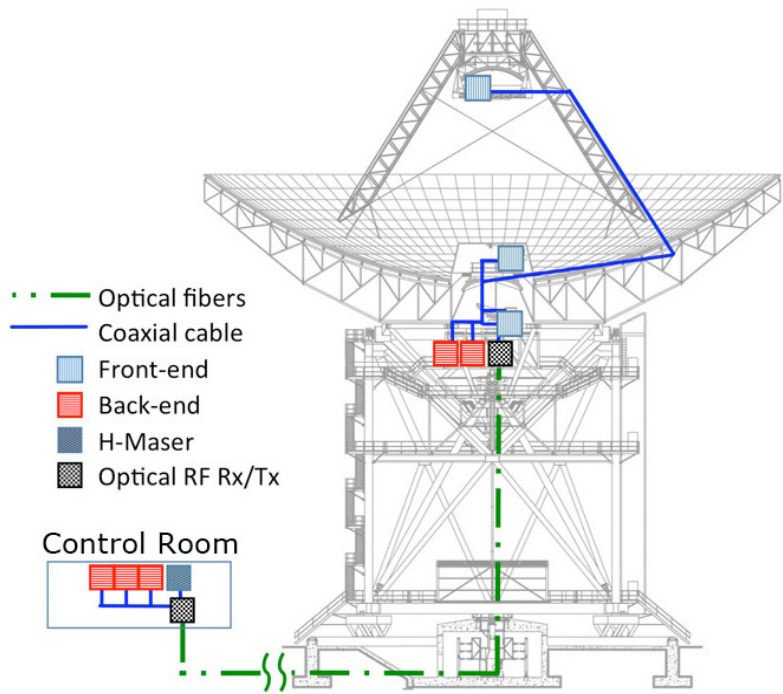

Fig. 5. Schematic view of the signal transfer lines from the front-ends to the back-ends.

In order to connect the radio telescope to the Internet, mainly to transfer data during real-time VLBI observations, SRT will make use of a 10-Gbps link infrastructure, currently under construction.

### 2.2. *Microwave receivers and back-ends*

Notwithstanding the telescope's enormous collecting area and its advanced systems, able to ensure high efficiency at all the operating frequencies and over the whole elevation range, state-of-the-art microwave receivers and digital and analog back-ends are necessary in order to perform breakthrough astronomical discoveries. The microwave receiver guides, amplifies, filters and down-converts the incoming astronomical signal reflected by the mirrors. It consists of a cascade of different microwave passive and active components, some of them cooled to cryogenic temperatures. The receiver mainly defines the frequency bandwidth, the antenna illumination, the receiver noise temperature and the cross-polar purity. The radio receiver group of INAF is responsible for the design of the whole receiver chain, both from an electronic and mechanical point of view, for the assembly and testing and finally for the integration in the radio telescope.

The first-light of SRT was obtained using the following three receivers: the L- and P-band dual-frequency coaxial receiver for primary-focus operations (Valente *et al.*, 2010), the C-band monofeed receiver designed for the BWG *layout I* (Nesti *et al.*, 2010; Orfei *et al.*, 2011; Poloni, 2010; Peverini *et al.*, 2011) and the K-band multi-feed receiver (seven corrugated feeds) for the secondary focus (Orfei *et al.*, 2010).

For the very first astronomical tests, the C-band receiver was installed in the Gregorian focus with a cone section added in front of the feed-horn to match the different F/D numbers. This choice was made to avoid possible BWG mirror misalignments in the first pointing tests of the antenna.

Table 3 describes the main technical characteristics of the receivers currently installed, together with those foreseen to be completed in the next three years. For each receiver the table lists the observable sky frequency bandwidth, the focal position, the number of simultaneous beams in the sky (specifying the number of polarizations), an estimation of the expected antenna gain and of the system noise temperature when the antenna points to the zenith. Finally, the last column indicates the status of the receiver.

All the multi-feed receivers (S-, K- and Q-band) are equipped with a built-in mechanical rotator, whose aim is to prevent sky field rotation while the telescope is tracking.

Table 3. Microwave receivers installed and under construction for the SRT.

| Receiver | Freq range [GHz] | Focal position | Pixels × polarizations | Expected antenna gain [K/Jy] | Expected system temperature at zenith [K] | Status |
|---|---|---|---|---|---|---|
| L- and P- band coaxial feed | 0.305–0.410 | F1 | 1 × 2 | 0.47–0.59 | 50–80 | Commissioned |
|  | 1.3–1.8 |  | 1 × 2 | 0.50–0.60 | 17–23 |  |
| C-band mono-feed | 5.7–7.7 | F3 | 1 × 2 | 0.64–0.70 | 24–28 | Commissioned |
| K-band multi-feed | 18–26 | F2 | 7 × 2 | 0.60–0.66 | 40–70 | Commissioned |
| S-band multi-feed | 2.3–4.3 | F1 | 5 × 2 | 0.76 | 54 | Under construction |
| C-band (low) mono-feed | 4.2–5.6 | F4 | 1 × 2 | 0.62–0.70 | 30–35 | Under construction |
| X- and Ka-band coaxial feed | 8.2–8.6 | F1 | 1 × 1 | 0.64 | 120 | Under testing |
|  | 31.8–32.3 |  | 1 × 1 | 0.57 | 190 |  |
| Q-band multi-feed | 33–50 | F2 | 19 × 2 | 0.45–0.56 | 45–120 | Under construction |
| W-band mono-feed | 84–116 | F2 | 1 × 1 | 0.34[a] | 115 | Under refurbishment |

[a]With a surface accuracy of 178 micron and the metrological systems in operation.

The W-band receiver, which was decommissioned by the Plateau de Bure radio telescopes operated by IRAM, is mainly aimed at testing the SRT active surface, and at acquiring know-how in the 3-mm band. However, even though it is a single-pixel, one-linear-polarization and narrow-instantaneous-bandwidth (600 MHz) receiver, its installation on the SRT could play a relevant role in the mm-VLBI network, significantly extending its baselines.

The front-end outputs are connected to the back-ends either by coaxial cables (in case they are placed within the radio-telescope, in particular the Total Power and XARCOS back-ends) or via optical fiber in order to reach the distant control room. Back-ends represent key elements to the success of the observations. They can be either general-purpose or tailored to specific scientific activities. All the back-ends listed in Table 4 were installed and are now either commissioned or under test at the telescope (Melis *et al.*, 2014).

### 2.3. *Optical system and active surface*

The SRT optical system is based on a quasi-Gregorian profile with shaping applied to both the primary and the secondary surfaces. The present geometry results from a trade-off between two goals: minimizing the overall system noise temperature, mainly due to spill over, and reducing the standing wave pattern (due to internal reflections between the primary and the secondary mirrors and thus detrimental to wideband spectroscopic observations) without excessively sacrificing the Field of View available from the Gregorian focus (Cortès-Medellin, 2002). The shaped parabola–ellipse pair, in fact, provides a wider focal plane than fully shaped reflectors; the FOV's radius is 20 $\lambda$ at 10% of aperture efficiency loss. The design placed more emphasis on reducing the standing wave than on maximizing the aperture efficiency.

Table 4. Back-ends available at SRT. In brackets are given the applications currently under testing or refinement and planned to be fully operative within 2016.

| Back-end | Main features and usage | Inputs | Max input IF band (GHz) | Integrat. time | Max spectral channels | Min spectral resolution (Hz) |
|---|---|---|---|---|---|---|
| Total Power | Bandwidth selectable Attenuation selectable IF distributor Continuum | 14 | 0.1–2.1 | 1–1000 ms | — | — |
| DFB | ADCs, 10 bit resolution 10 bit DSP Pulsar (Spectro-polarimetry) | 4 | 0–1.024 | 0.1 ms–4 s | 8192 | ∼1000 |
| XARCOS | Spectro-polarimetry | 14 | 0.125 | 10 s | 2048 | ∼250 |
| DBBC | 8 bit DSP VLBI (RFI monitoring) (Spectroscopy) | 4 | 0.512 | 0.1–1 s | 4096 | 125,000 |
| ROACH | ADCs, 8 bit resolution LEAP - Pulsar, array (Pulsar, single dish) (Spectroscopy) | 2 | 0.512 | 1 s | 8192 | 62,500 |

One of the most innovative features of the SRT is the active surface that consists of 1116 electromechanical actuators mounted in the backup structure, beneath the primary mirror panel corners, and distributed along radial lines. Its first aim is to reshape the mirror to compensate for the repeatable deformations due to gravity. Exploiting advanced real-time measurements, it will also correct for wind and thermal effects (Orfei *et al.*, 2004). Each actuator moves either upward or downward at the corners of four adjacent panels, in the direction normal to the local surface, with a maximum stroke of ±15 mm. Such a large stroke was required in order to achieve the second aim, i.e. to modify the shaped profile back to a surface that is parabolic enough to increase the maximum operating frequency observable from the primary focus (Bolli *et al.*, 2014). A further application of the active surface is also to recover the manufacturing deformations of the secondary mirror (Bolli *et al.*, 2003).

In order to perform high frequency observations the surface accuracy must be around 150 $\mu$m. The contractual requirement for the alignment of the primary reflector was 500 $\mu$m with a "goal" of 300 $\mu$m. Successive accurate photogrammetry measurement campaigns have shown a lower limit in the total RMS of the primary mirror equal to 290 $\mu$m at 45° of elevation (Süss *et al.*, 2012). These measurements, carried out in six different elevation positions, allowed the production of a look-up table, used to correct the gravity deformation with the actuators in an "open loop" configuration. In order to further improve the measurements of the primary mirror profile, microwave holography measurements will be implemented by using the hardware setup successfully tested at the 32-m Medicina radio telescope (Serra *et al.*, 2012).

Moreover, several solutions are being studied and implemented so as to perform real-time measurements of the reflector deformations and the misalignments errors in the subreflector position, with the final aim of correcting them in a "closed-loop" strategy (Pisanu *et al.*, 2012; Prestage *et al.*, 2004):

(i) improved FEM analysis — to understand the effects due to gravity, temperature and wind on the behavior of the mechanical structure of the antenna, and to guide the optimized design of the sensor systems;
(ii) inclinometers — to monitor the status of the track, the inclination of the azimuth axis and the variations of the alidade deformation as a function of the temperature;
(iii) position-sensing devices — to optically monitor the lateral shift of the secondary mirror;
(iv) optoelectronic linear sensors — to measure the deformations of the primary mirror;
(v) optical fiber rangefinders — to measure the deformations of the legs of the quadrupod in order to control the sub-reflector position;
(vi) thermal sensors — to map the distribution of temperature in the most sensitive areas of the radio telescope and predict the deformations induced by thermal gradients.

## 2.4. *Antenna control software*

The SRT consists of a large number of modules and devices to be managed and controlled in a timely and precise fashion. The many observing modes available must be commanded and harmonized with the data acquisition and the recording of timestamps and housekeeping data. The overall control software performing all of these actions is called *Nuraghe*, a software package exceeding half a million lines of code. It runs in the framework of the ALMA Common Software (ACS, a distributed-object framework in turn based on CORBA), which provides a general and common interface for high-level software, hardware and other software parts above the operating system.

For single-dish operations, various modes are supported such as raster scan, On-The-Fly (OTF) cross-scan and mapping, sky dipping, position switching and focusing. The system is also able to operate all the three Italian antennas under a common interface (user interface and scheduling system), as well as to accommodate external applications such as VLBI or pulsar observations.

All data coming from the integrated back-ends are stored in a FITS file. The current release of Nuraghe (Orlati *et al.*, 2012), which is written in C++ and Python, enables the operator to control most of the subsystems — e.g. the antenna mount, the receiver chains, the back-ends, the active surface and the minor servo (Buttu *et al.*, 2012).

### 2.5. *Auxiliary instrumentation*

We now discuss several associated ancillary activities such as radio spectrum monitoring, weather parameters measurement and time-frequency standard distribution, as they are fundamental aspects for the optimal operation of the SRT.

A mobile laboratory is available at the site to carry out dedicated measuring campaigns to monitor and identify Radio Frequency Interference (RFI) in the frequency range between 300 MHz and 40 GHz (Bolli *et al.*, 2013). Special attention is paid to the frequency bands observed by the first-light receivers and to those allocated by the Italian Frequency Allocation Plan to the Radio Astronomical Service (see Sec. 3.3).

A rather sophisticated ground-based set of weather measuring instruments are deployed at the SRT site. Besides conventional sensors — ambient thermometer, hygrometer, wind speed/orientation measurer and barometer — a multi-frequency radiometer is employed. It generates real-time estimates of different microwave brightness temperatures, opacity, zenith water vapor and liquid content. The measurements conducted so far show a 40–45% probability of finding an integrated water vapor column density of less than 10 mm in winter time. The absence of cloud cover can be found 50% of the time in winter and 80% during summer; typical liquid water values range between 0.2 and 0.7 mm.

In winter, the opacity at 22 GHz is lower than 0.15 Np for 90% of the time (40% during summer). At higher frequencies, in the 3-mm band, the winter opacity is lower than 0.15 Np for 35% of the time.

The Time and Frequency laboratory consists of an active hydrogen maser producing the reference signals for the receiver local oscillators, the 1 PPS (Pulse per second) timing for data acquisition and all the related equipments. A dedicated GPS receiver derives the local clock offset (as a difference from UTC GPS) for VLBI and provides the time to the antenna for its pointing via an IRIG-B generator (Ambrosini *et al.*, 2011b).

## 3. Performance Test

### 3.1. *The first radio source*

SRT detected its first radio source in Summer 2012. The observation was performed with the C-band receiver, installed in the Gregorian focus at the time. The main uncertainties at that time were the encoder alignments of the two main movement axes. Using the Moon, a wide and bright target, as a first reference radio source and moving the antenna in steps of a half beam-size in a cross-scan, we measured the following encoder offsets: $+1.3°$ in azimuth and $-0.5°$ in elevation. The successive OTF scans were performed across 3C218, chosen because, being not too far from the Moon on that day, it allowed us to use the just-measured "local" offsets, in the absence of a pointing model. The antenna response is illustrated in Fig. 6; the noticeable pattern asymmetry is due to the employment of fixed optics, observing far from the elevation of their mechanical alignment $(45°)$.

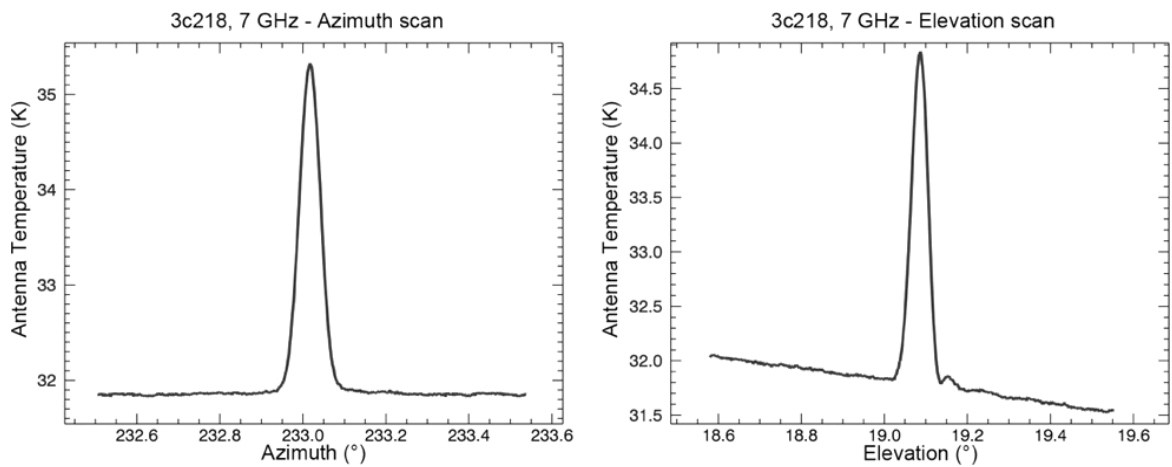

Fig. 6. The first observed radio source, 2012 August 8.

### 3.2. *Pointing accuracy*

The SRT software was designed to manage a complete and independent model for each receiver and each focus. Three models were thus processed, initially considering $11.3'$, $2.7'$ and $0.8'$ as reference Half-Power Beam-Widths (HPBWs) for the L-band, C-band and K-band receivers, respectively.

To achieve a preliminary pointing model, we applied the encoder offsets obtained during the first radio source observation and produced many raw maps over bright and compact sources in order to measure the displacement of the optical axis with respect to the ideal one. This procedure permitted the creation of a basic set of parameters, which was used as a basis for the pointing models for all the three foci.

As a next step to improve the pointing model accuracy, we performed several series of spot observations (OTF cross-scans along the Horizontal frame) of selected pointing calibrators, each time measuring the pointing offsets. In order to minimize thermal effects on the structure we usually performed the observation during the night, in any case avoiding the sunrise. Once the whole azimuth/elevation plane in the sky was satisfactorily sampled, the pointing model polynomial was fit to the dataset (Maneri & Gawronski, 2002; Guiar *et al.*, 1986). This process required several sessions to converge to an acceptable solution, for which the model residuals are required to be lower than one tenth of the beam size.

The RMS residuals after the fit are presented in Table 5 and show considerably less than one tenth of the beam size up to 23 GHz. The pointing performance at higher frequencies (the beam dimension will be around 12 arcsec at 3 mm) will take advantage of metrological systems (as discussed in Sec. 2.3).

### 3.3. *RFI*

Dedicated campaigns were carried out, both with the mobile laboratory and through the telescope, in order to detect and identify the interfering signals. Generally speaking, the RFI environment was found to be fairly quiet in the C-band (especially above 6400 MHz) and the K-band (almost everywhere in the band), while the P- and L- bands appeared more polluted. Figure 7 shows the spectral acquisitions performed with the various SRT receivers. These scans were carried out spanning the whole azimuth range, repeating such 360° circles for different elevations (every 3° above the horizon). These charts were obtained in "max hold" mode: all the instantaneous acquisitions are overplot and, for each frequency bin, the maximum recorded value is displayed, thus the diagrams represent a sort of "worst case scenario" of the signals detected in the above specified az–el ranges.

The L- and P-band environment is the most "polluted" at the site. Under these conditions, in order to find a suitable frequency configuration for the Total Power back-end, whose narrowest built-in band is 300 MHz, we installed external tuneable (5% of the central frequency) band-pass filters. For L-band acquisitions, an effective compromise was found and we successfully observed between 1696.5 and 1715.0 MHz. No solution, allowing reliable and repeatable measurements in the P-band, was found. Almost all the interfering signals were identified to be self-generated by the apparatus installed in the telescope. Among these devices we can mention the VOIP phones in Alidade and elevation Equipment Rooms, the XARCOS back-end, the encoders of the PFP, and some devices of the control electronics of the K-band receiver.

As concerns the C-band receiver, the general panorama was quite satisfactory; we discovered the presence of only a few interfering signals, mainly concentrated in the lower part of the receiver bandwidth. The signals at 5900 and 6016 MHz have been identified to be self-generated. The former is produced by one of the local oscillators of the multifeed system, the latter by the device that guarantees Internet connection to the station through a satellite link. The RFI signals coming from external sources were all due to digital links, and they appeared to be much attenuated when observing southward.

The K-band turned out to be almost free from RFI. All the identified polluting signals came from external sources were related to fixed links for mobile operators. These signals can be practically neglected by avoiding known azimuth and elevation positions. The bands allocated to radio astronomical service (22–22.5 and 23.6–24.0 GHz) were confirmed to be fully usable.

Table 5. Pointing model residuals and beam size for the different receivers.

| Rec. | Observed center freq. [GHz] | Pointing model residuals [arcsec] | | Beam size [arcmin] |
|---|---|---|---|---|
| | | $\sigma_{az}$ | $\sigma_{el}$ | |
| L | 1.705 | 7 | 7 | 11.12 |
| C | 7.35 | 10.8 | 7.2 | 2.58 |
| K | 23 | 3.9 | 3.2 | 0.805 |

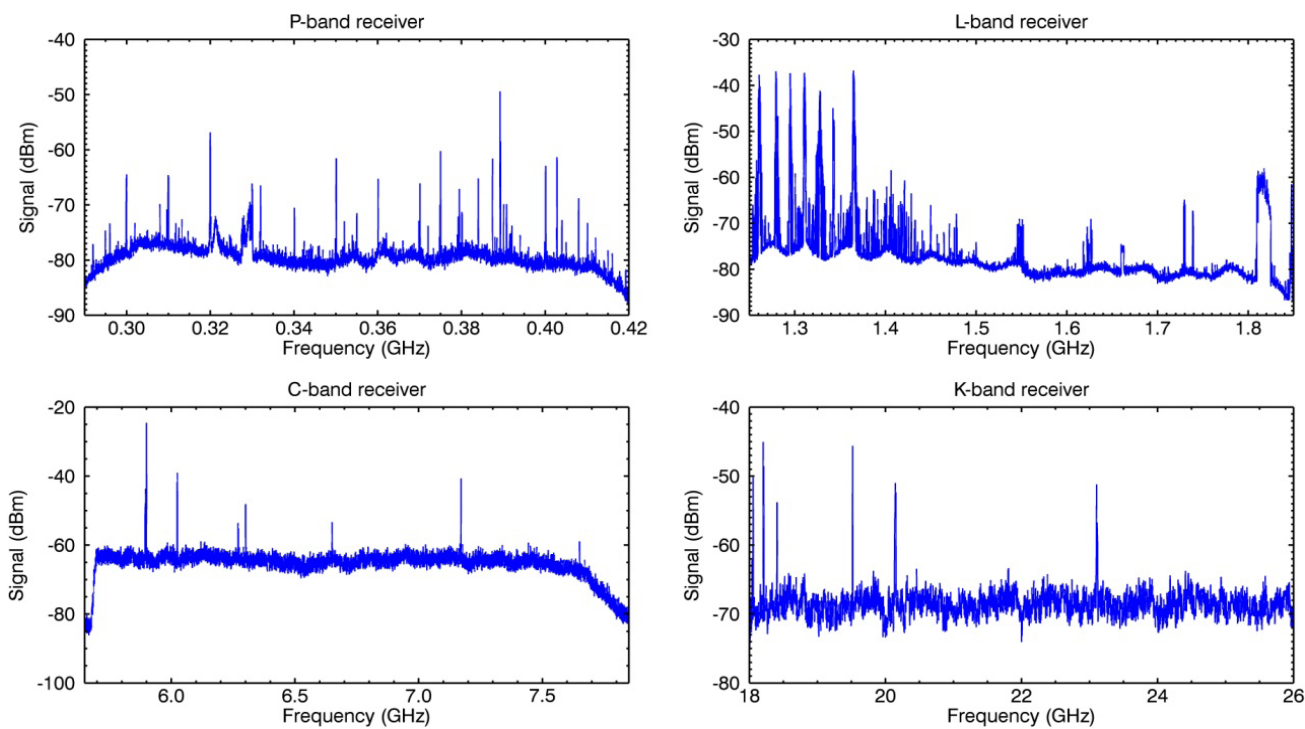

Fig. 7. Spectral plots illustrating the RFI affecting the various receiver bands. These "max hold" plots show all the signals received while observing in the whole azimuth range, for several elevations above the horizon.

We emphasize that several interfering signals turned out to actually be "self-RFI", i.e. they were produced by devices installed at the SRT site. Since most of those devices are going to be moved into a properly shielded room in a very near future, we expect that the RFI environment will greatly improve and we thus do not consider the acquisitions performed during the commissioning to be representative of the final site conditions. It must also be stressed, for the sake of the future users of the SRT, that spectral back-ends — and specific mitigation techniques — are being developed in order to cope with RFI and are expected to be operative within 2016 (see Table 4).

### 3.4. *BWG focus commissioning*

#### 3.4.1. *System temperature*

The C-band receiver measured system temperature versus the elevation position is shown for both polarizations (LCP and RCP) in Fig. 8. Values were obtained for every 10° of elevation in the 6.7–7.7 GHz sub-band. The atmosphere opacity was estimated by performing a skydip scan ($\tau_0 = 0.014$).

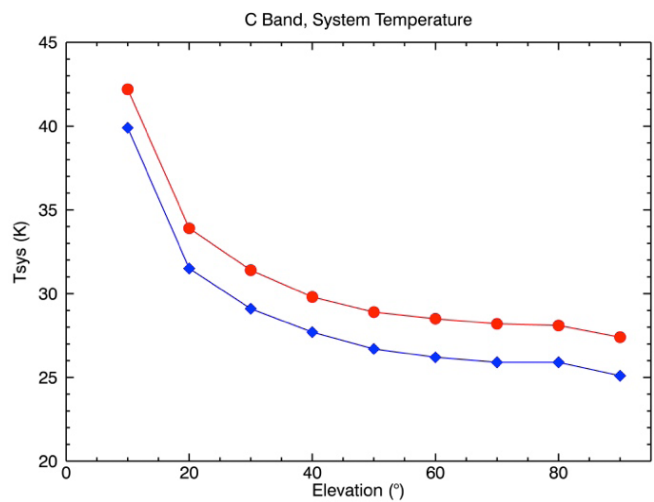

Fig. 8. Tsys measured in the band 6.7–7.7 GHz. Circles: LCP. Diamonds: RCP.

The single expected contributions to the system noise temperature are listed in Table 6. The sum of these elements (23.7 K, 22.0 K) is comparable to the obtained measures, 27 and 25 K, respectively for LCP and RCP. The 2-K difference between the experimental curves is consistent with the different receiver temperatures ($T_{\rm rx}$), whereas the slight differences between theoretical and experimental values can be explained with a higher-than-predicted

Table 6. Contributions to the Tsys in the 6.7–7.7 GHz band.

|  | Contribution | Notes |
| --- | --- | --- |
| $T_{\rm rx}$ | LCP 8.5 K | Lab measurements |
|  | RCP 6.8 K |  |
| Gregorian cover | 1.4 K | Observations |
| Atmosphere + CMB | 4.1 K | $\tau = 0.014$ at zenith |
| Ground spill-over | 2.7 K | Simulation |
| BWG spill-over | 7.0 K | Simulation |

spill-over temperature (mainly coming from the BWG mirrors).

Table 6 includes an extra-noise term caused by the cover of the Gregorian room, a 1-mm Teflon film 1.5 m in diameter, which protects the apparatus from the external environment. The film trade name is Virgin PTFE, and the refraction index (not provided as a function of frequency) declared by the manufacturer is 1.35, yielding a reflection amounting to 4%. We measured the Tsys increase by temporarily removing the cover; results are summarized in Table 7, where the average of the recorded increments is shown for different frequency bands. As shown in Table 7, such a contribution appears in the C-band measurements, but it is definitively more significant in K-band.

### 3.4.2. *Gain curve*

The efficiency measurement and the gain curves were produced exploiting the continuum back-end, performing cross-scans in the Horizontal frame on bright, point-like calibrators (Table 8). Atmospheric opacity was extrapolated from skydip scans.

This campaign was repeated in two phases, reflecting two different telescope configurations. During the first phase, the main mirrors were fixed to the reference mechanical alignment. The second session was carried out when the active surface and the sub-reflector were aligned according to photogrammetry, in order to compensate for the gravitational deformations that vary with elevation.

The antenna gain in the 7.0–7.7 GHz band is shown in Fig. 9 for the fixed optics. The plot compares the theoretical values (triangles) with the experimental measurements for the LCP channel (circles). The expected curve was obtained taking into account both the receiver and the antenna efficiency parameters, including the surface efficiency computed from the RMS of the mirrors, in the hypothesis that the optics were aligned at all the elevations (not achievable in real measurements, as the subreflector was not tracking). The trend of the measured curve reflects the expected one, however a gain offset (about 10%–15%) is evident at all elevations — including 45°, where the antenna optics were supposed to be effectively aligned.

Table 7. Tsys contribution by the Gregorian room cover.

| Cover, Tsys contribution (K) | | | | |
| --- | --- | --- | --- | --- |
| 6.7–7.7 GHz | 18–20 GHz | 20–22 GHz | 22–24 GHz | 24–26 GHz |
| 1.4 | 11.2 | 9.2 | 8.3 | 9.8 |

Table 8. List of the flux density calibrators used for C-band measurements (Ott, 1994).

| Source | RA J2000 | Dec J2000 | Flux density at 7.35 GHz [Jy] |
| --- | --- | --- | --- |
| 3c48 | 01:37:41.2971 | +33:09:35.118 | 3.7975 |
| 3c147 | 05:42:36.1379 | +49:51:07.234 | 5.2911 |
| 3c286 | 13:31:08.2881 | +30:30:32.960 | 5.7071 |
| 3c309.1 | 14:59:07.578 | +71:40:19.850 | 2.3433 |
| 3c295 | 14:11:20.6477 | +52:12:09.141 | 4.0588 |
| 3c161 | 06:27:10.096 | −05:53:04.72 | 4.3614 |

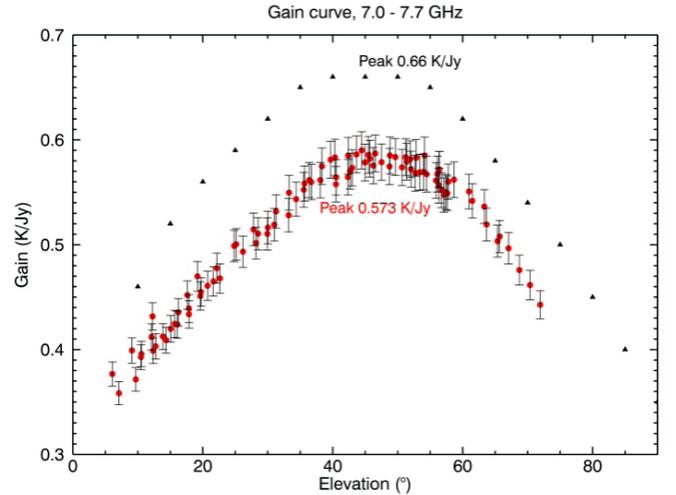

Fig. 9. Gain with mechanically aligned mirrors: measurements (red circles) and expected values (triangles).

The plots of Fig. 10 show the BWG receiver gain curves for both the circular polarizations, obtained with active surface and subreflector tracking. The flatness of the curves along all the elevation range demonstrates a great improvement due to the compensation for structural deformations; on the other hand, the peak is slightly below expectation (0.61 instead of 0.66 K/Jy). We might then conclude that the optical alignment of the main and secondary mirrors is quite satisfactory for this wavelength, and that the employment of the active surface is able to deliver an almost constant gain.

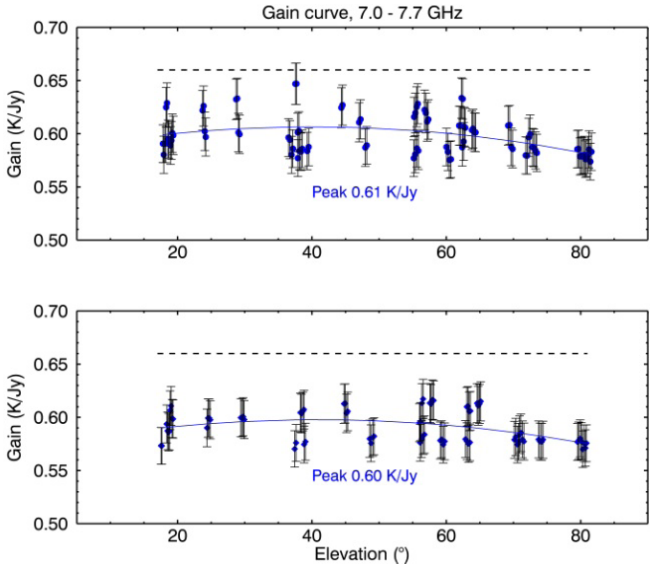

Fig. 10. C-band measured gain curves with the active surface and the sub-reflector movement operating: LCP (top), RCP (bottom). Dashed lines indicate the expected values in an ideal scenario.

### 3.4.3. *Beam size*

Figure 11 shows the beam deformation ($\varepsilon$) with respect to the estimated HPBW (beam size, $\theta_e$), for both telescope configurations (fixed and active surface). The beam size is a reliable indication of how the optics of the telescope reflect the theoretical design; accordingly, the deformation of the beam along the elevation span could be a symptom of a flawed telescope structure. The beam deformation is computed through Eq. (1), where $\theta_{\rm az}$ and $\theta_{\rm el}$ are the beam sizes measured along the azimuth and elevation axis, respectively. The comparison of the two curves led us to the same conclusion inferred through the gain curves: the efficiency reduction observed with the fixed alignment is almost completely recovered using the active surface and the tracking subreflector.

$$\varepsilon = \sqrt{(\theta_{\rm az} - \theta_e)^2 + (\theta_{\rm el} - \theta_e)^2}. \quad (1)$$

### 3.5. *Gregorian focus commissioning*

The characterization of the K-band receiver, located in the Gregorian focus, was obviously complicated by the impact of weather conditions on radiation at these wavelengths. It was thus decided, so as to achieve very accurate and effective measurements, to limit the commissioning activities to the time intervals showing low and stable opacity. During these clear-sky periods the opacity at the site turned out to be particularly favorable.

#### 3.5.1. *System temperature*

The K-band receiver Tsys measurements were performed following the same procedures employed for the BWG receiver, for both the polarizations, repeating the procedure for each 2-GHz sub-band. Figure 12 shows the LCP values for the central feed.

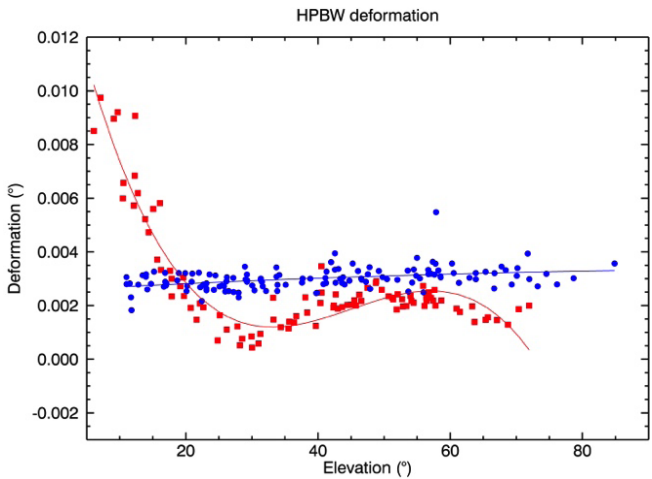

Fig. 11. C-band beam deformation. Squares: with fixed optics. Circles: with active surface and tracking subreflector ($\sigma = 1.45$ arcsec).

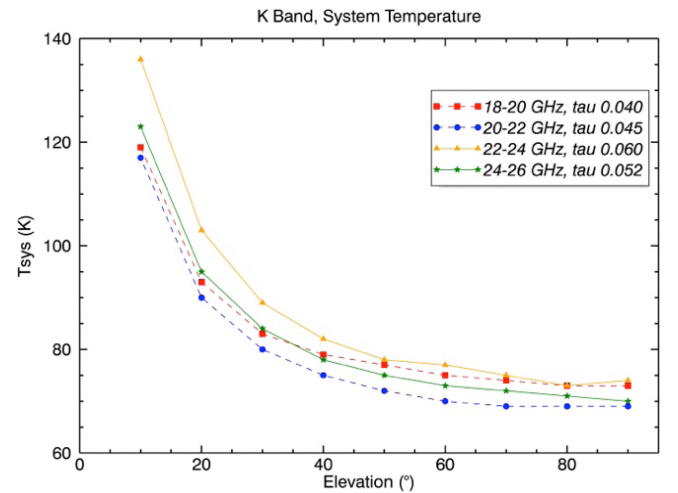

Fig. 12. Tsys versus elevation measured in 2-GHz sub-bands, LCP of central feed. The zenith opacity values are also reported.

Table 9. Tsys contributions in all the receivers sub-bands, measured with the central feed.

|  | 18–20 GHz | 20–22 GHz | 22–24 GHz | 24–26 GHz | Notes |
| --- | --- | --- | --- | --- | --- |
| $T_{\rm rx}$ LCP | 23.7 K | 24.4 K | 29.7 K | 33.9 K | Lab meas. |
| $T_{\rm rx}$ RCP | 21.6 K | 18.0 K | 22.7 K | 27.1 K | |
| Gregorian cover | 11.2 K | 9.2 K | 8.3 K | 9.8 K | Observation |
| Atmosph. + CMB ($\tau$) | 11.4 K | 12.7 K | 16.8 K | 14.7 K | El = 90° |
|  | (0.040) | (0.045) | (0.060) | (0.052) | |
| Ground spill-over | 2.3 K | 0.6 K | 0.6 K | 0 K | Simulation |

The atmospheric contribution was evaluated by means of skydip scans, whose results were compared to the radiometer measurements. Table 9 lists the main Tsys contributions.

Even taking into account the extra system temperature contribution due to the cover film, we find that the measured values do not match expectations. The sum of these contributions reaches 49 K, quite distant from the observed value of 73 K if we refer to the 18–20 GHz band. The cause of such a difference between measured and theoretical was not completely understood and therefore is still under investigation.

### 3.5.2. *Gain curve*

The list of calibrators observed during the efficiency measurements is given in Table 10. Each calibrator was sampled after performing preliminary cross-scans to focus the telescope and achieve an optimized pointing. The atmospheric opacity was again estimated by means of skydip acquisitions.

A preliminary gain curve, achieved in the 22–24 GHz band, was acquired with the fixed optics configuration. The experimental and the theoretical results, in Fig. 13, are comparable. Measurements show lower values with respect to simulations, as the latter were computed taking into account constantly aligned — yet non-active — optics, while measurements were carried out keeping the mirrors in their reference positions, i.e. the positions obtained with their mechanical alignment at 45° of elevation. In this reference position, where the optics alignment should be ideal, a slight loss in the peak gain is still noticeable (0.52 versus 0.55 K/Jy). This can be explained as a residual inaccuracy in the mechanical alignment.

Table 10. Flux density calibrators, used for the K-band gain curve (Ott, 1994).

| Source | RA J2000 | Dec J2000 | Flux density at 22.35 GHz (Jy) |
| --- | --- | --- | --- |
| 3c48 | 01:37:41.2971 | +33:09:35.118 | 1.2057 |
| 3c147 | 05:42:36.1379 | +49:51:07.234 | 1.7655 |
| 3c286 | 13:31:08.2881 | +30:30:32.960 | 2.4330 |
| NGC7027 | 21:07:01.593 | +42:14:10.18 | 5.3890 |

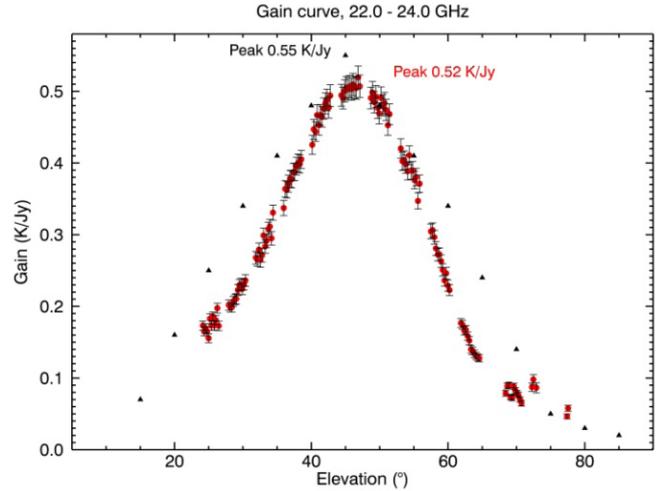

Fig. 13. Opacity-corrected gain curves for the K-band receiver, with fixed optics. Theoretical values (triangles) were computed considering aligned mirrors at each elevation. Measurements (circles) were acquired with both the mirrors fixed in the mechanically aligned positions.

Similarly, the gain curves for both LCP and RCP were measured enabling the active surface and the subreflector tracking. We expected the curve to be flat at 0.66 K/Jy, taking into account the RMS of the surfaces of mirrors, the alignment of M1 — estimated with photogrammetry — and other involved parameters such as: accuracy of the alignment of the subreflector panels, errors in the measurements on panels, gravitational, thermal and wind effects on panels, positioning accuracy of the actuators. The plots in Fig. 14 clearly show that, in terms of peak gain, the data match the predictions. On the other hand, even if the overall telescope efficiency benefits from the active surface and the tracking of the subreflector, there is still a substantial decrease in the efficiency below 45° of elevation. This excess of gain loss at lower elevations can easily be translated in a surface accuracy of about 500 micron (El = 20°). Our investigations suggest that gravity deformation measurements were not sufficiently accurate at low elevations. In view of the installation of higher frequency receivers, this problem shall be solved; however, this will be done in the refinement phase by using the microwave holography technique.

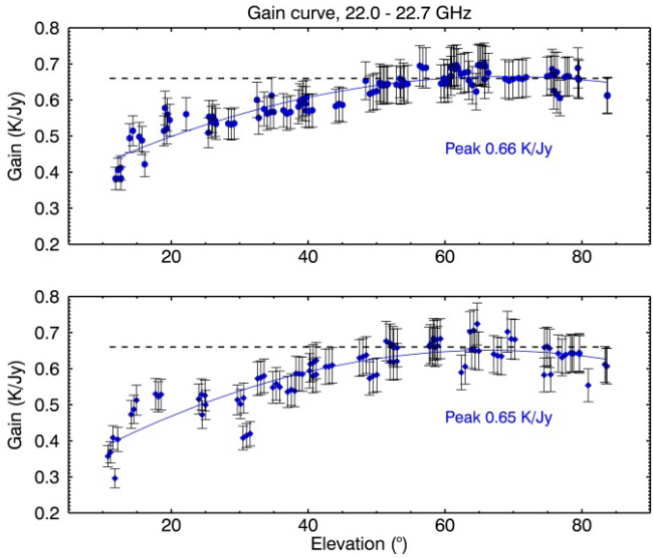

Fig. 14. Gain curve in the 22.0–22.7 GHz band. LCP (top) and RCP (bottom) for the central feed, acquired enabling the active surface and the subreflector tracking. Dashed lines represent the expected values in ideal conditions.

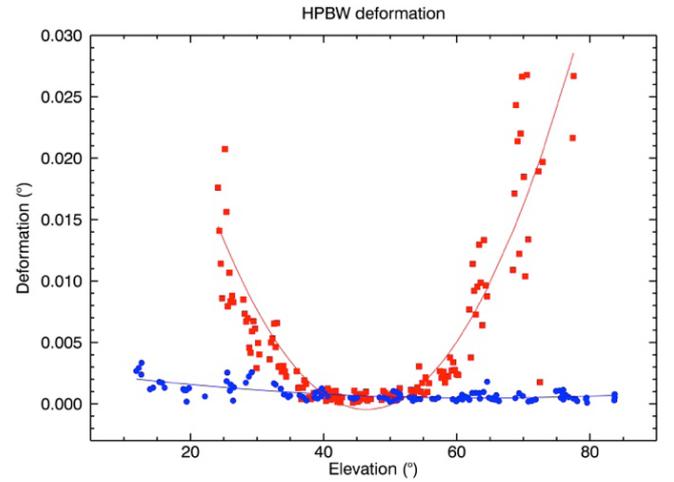

Fig. 15. Beam deformation, with respect to the ideal beam size, versus elevation. Squares: with fixed optics. Circles: with active surface and subreflector in tracking mode ($\sigma = 2.25$ arcsec).

### 3.5.3. *Beam size and bidimensional pattern*

Measurements of the beam size and pattern were performed alternatively enabling and disabling the active surface, in order to verify its effectiveness and to roughly evaluate the quality of the optical alignment. The outcome of these experiments was also meant to be exploited to draw further conclusions about the observed gain curves, and to prove whether there was a relation between potential alignment inaccuracies and the gain drop observed at low elevations.

Figure 15 compares the beam deformation $\varepsilon$ (Sec. 3.4.3, Eq. (1)), measuring the deviation from the nominal HPBW, observed in the two telescope configurations (fixed and active surface) by means of cross-scans on non-resolved sources. Figure 16, instead, shows two raw maps in arbitrary counts, acquired on one of these sources. They illustrate the antenna beam patterns, respectively obtained with the fixed and active mirrors, in the same elevation range.

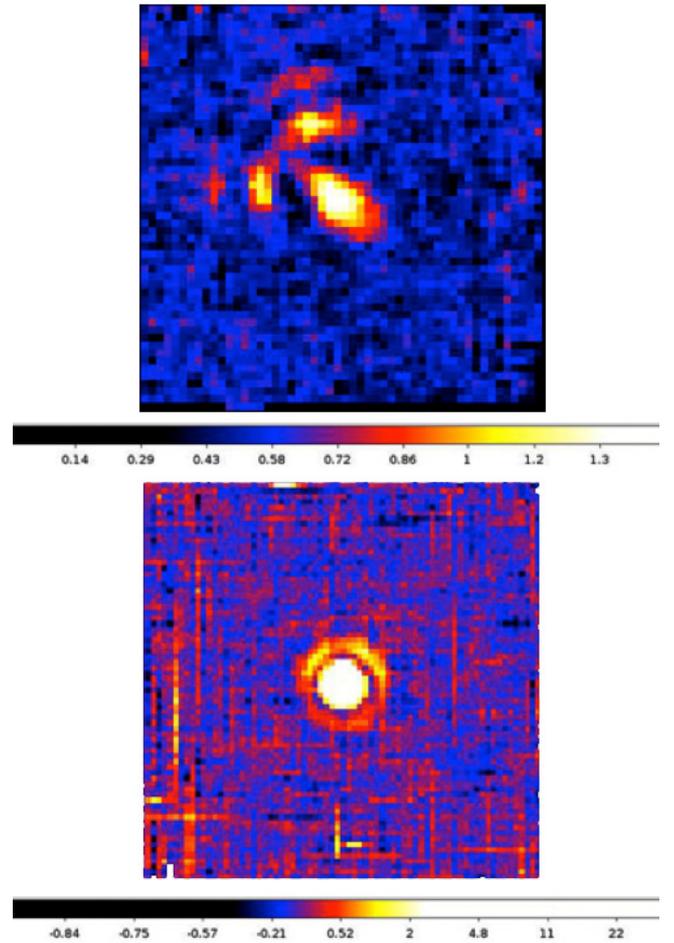

Fig. 16. Two az–el raw maps of the same point-like source. Fixed optics (top) and active surface plus tracking subreflector (bottom). Both for El > 60°.

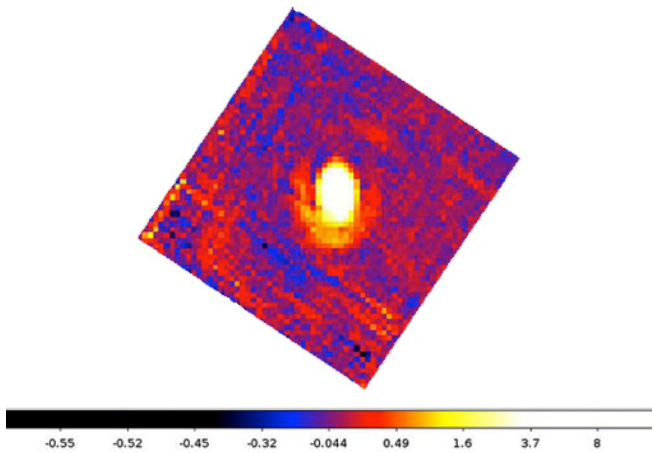

Fig. 17. Az–el raw map over a point-like source with active surface enabled and tracking subreflector (average elevation 22°).

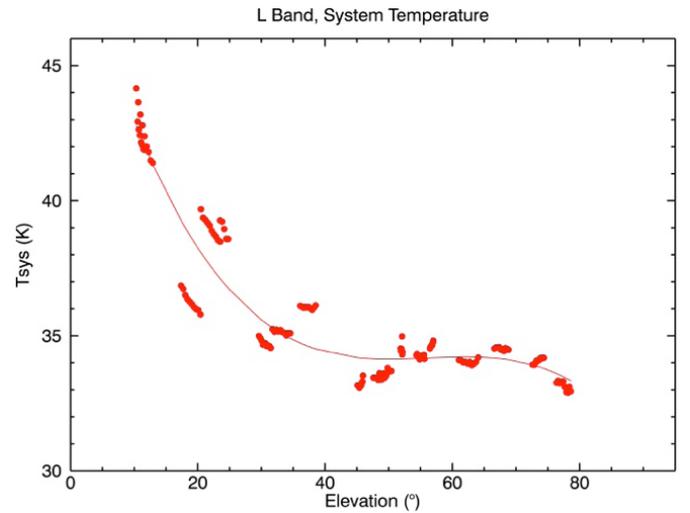

Fig. 18. System noise temperature versus elevation for the L-band primary-focus receiver (1.6965–1.7150 GHz).

The telescope in fixed optics configuration showed a beam size rapidly deforming as the pointed elevation deviated from the mechanical alignment elevation (45°). Astigmatism and coma lobes, revealed in Fig. 17 (top), explain such a distortion, at least for the upper range of elevation (El > 60°), that is compatible with the gain drop shown in Fig. 13. The pattern in Fig. 17 (top) shows also a distortion tilting at about 45° with respect to the az–el axes. This deformation is not aligned with the gravity vector, so it must be a contribution of several different ones. The compensation of gravity deformations obtained employing the active surface of the primary mirror, on the other hand, allowed us to obtain a particularly effective alignment for elevations above 45°. Figure 17 (bottom) demonstrates that most of the distortions of the antenna beam pattern had disappeared in this range. An asymmetry in the sidelobes is still visible, likely due to some residual squint.

Finally, the map in Fig. 18, taken at the average elevation of 22°, highlights a significant asymmetry and deformation, visible both in the main beam and in the first sidelobe. This deformation can be considered in agreement with the characteristics of the gain curve presented in Fig. 14, as it can be explained by the current status of the optics.

### 3.6. *Primary focus commissioning*

#### 3.6.1. *System temperature*

After the preliminary tests, the primary focus receiver turned out to have an issue in the microwave chain responsible for the signal phase shifting producing circular polarizations from the native linear ones. The problem affected the RCP only: the noise calibration signal was not detectable for this channel. The cause, although immediately identified, could not be fixed before the completion of the commissioning, thus no further measurements could be performed. For this reason, this paper discusses only the results for the LCP.

Table 11 presents the theoretical contributors leading to an estimated temperature of 27.0 K when the telescope is parked at zenith and up to 44.0 K when the elevation angle is 5°. Even considering all the caveats, e.g. the intrinsic inaccuracy in the prediction of the noise calibration signal level — the system temperature is higher than expected possibly due to an underestimation of the simulated spillover and/or other noise sources. Figure 18 for example provides hints about the influence of RFI on the measured Tsys values: it shows clustered measurements, in particular in the elevation range from 15° to 50°, creating "jumps" in the Tsys trend. This likely derives from the presence of variable RFI affecting the signal level.

#### 3.6.2. *Gain curve*

The L-band gain curve is given in Fig. 19. It shows slightly scattered measurements but, in this case, the gain value equal to 0.52 K/Jy is entirely in agreement with the expectations (0.50–0.55 K/Jy for the band under test).

Table 11. Tsys contributions (1.6965–1.7150 GHz).

|  |  | Contribution | Notes |
|---|---|---|---|
| $T_{\rm rx}$ | LCP | 13.0 K | Lab measurements |
| Atmosphere + CMB | El = 90° | 5.0 K | Observations and |
|  | El = 5° | 22.0 K | simulations |
| Ground spill-over |  | 9.0 K | Simulations |

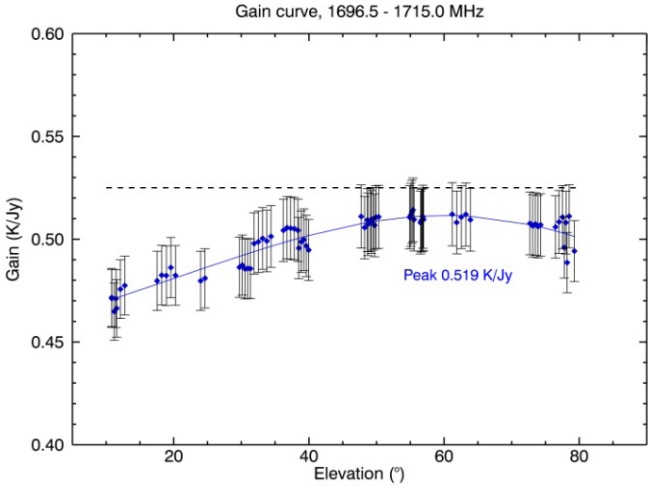

Fig. 19. Gain curve for the L-band receiver (1.6965–1.7150 GHz). The dashed line indicates the expected peak gain.

These measurements were performed using the active surface, yet only to re-shape the primary mirror in order to obtain a parabolic profile. Elevation-dependent corrections were not applied, as it is not necessary to compensate for gravity deformations at these low frequencies. The parabolic profile allowed us to increase the antenna gain with respect to the original quasi-Gregorian shaping. Using the latter, the antenna gain would rapidly decrease for frequencies above 1.3 GHz, reaching 0.4 K/Jy at 1.7 GHz. The re-shaping of the surface might allow us to profitably employ receivers up to a frequency of 22 GHz in the primary focus.

### 3.7. *Summary of the commissioning measurements and milestones*

The entire process of the telescope commissioning took almost 18 months, from mid 2012 to the end of 2013. Table 12 shows the schematic timeline of the major steps carried out, including examples of the technical advancement that has been taking place in parallel to the AV activities.

Table 13 finally summarizes the main telescope measured parameters.

Table 12. Main milestones of the commissioning and post-commissioning technical activities.

| Task description | Execution period |
|---|---|
| Antenna Software Interface tests | 2012 June |
| Gregorian Feed Rotator and M3 Rotator commissioning | 2012 July |
| Minor Servo Software integration and verification | 2012 July–2013 September |
| Gregorian focus receiver integration | 2012 July |
| First radio source with the C-band receiver | 2012 August |
| Gregorian Focus commissioning | 2012 August–2012 October |
| Primary Focus Positioner and Subreflector commissioning | 2012 August–2013 September |
| BWG Focus commissioning | 2012 October–2012 December |
| Primary Focus commissioning | 2013 May–2013 July |
| Active Surface Software integration and verification | 2013 January–2013 September |
| Inauguration | 2013 September 30 |
| Installation of back-ends (successive to the Total Power back-end) | 2013 October–2014 December |
| Upgrade of the Nuraghe control system to comply with the AV needs | 2015 June |

Table 13. Summary of the telescope performance measurements for the first-light receivers. Tsys and peak gain are measurable with an accuracy of 5% due to the intrinsic uncertainty in the temperature of the noise calibration signal.

| Rec. | Observed center freq. [GHz] | Beam size [arcmin] | Tsys, El = 90° [K] (±5%) | | Tsys, El = 30° [K] (±5%) | | Peak gain [K/Jy] (±5%) | |
|---|---|---|---|---|---|---|---|---|
| | | | LCP | RCP | LCP | RCP | LCP | RCP |
| L | 1.705 | 11.12 ± 0.11 | 34 | n.a. | 35 | n.a. | 0.52 (44.6%) | n.a. |
| C | 7.35 | 2.58 ± 0.02 | 27 | 25 | 31 | 29 | 0.61 (52.4%) | 0.60 (51.5%) |
| K | 23 | 0.805 ± 0.04 | 74 | 77 | 89 | 92 | 0.66 (56.6%) | 0.65 (55.8%) |

## 4. Conclusions

During the technical commissioning phase, all the devices foreseen for the telescope early activities were successfully tested. The SRT overall performance was proven to be close enough to expectations. Three receivers (L-, C- and K-band) and a Total Power back-end were fully characterized. The new antenna control system was continuously improved and updated during the activities, allowing for the execution of test observations in the most common single-dish modes. The capabilities of the primary reflector active surface and of the tracking subreflector were widely demonstrated, as the measurements performed enabling these devices turned out to be almost elevation-independent. Refinements regarding the optics are still needed only for K-band observations; further improvements, also in anticipation of the installation of even higher frequency receivers (up to 100 GHz), will be achieved utilizing microwave holography and metrology techniques, at present under investigation and testing.

As concerns the environmental conditions of the telescope site, our experiments confirmed that the location meets the requirements for high-frequency observations. The occurrence of RFI — aggravated by the local presence of temporarily unshielded apparatuses — was found to be an impediment only for P-band observations; the C band was mostly employable and the K band was confirmed to be particularly unpolluted — especially above 20 GHz.

The telescope was hence made available to the Astronomical Validation team, in charge of assessing its scientific potentials, while proceeding with the installation and testing of additional devices — such as digital spectrometers — in view of the shortcoming opening of the SRT to the worldwide community.

## Acknowledgments

The SRT project was funded mainly by the Italian Ministry of Education, University and Research, joined by ASI and the Sardinian Regional Government, the latter supporting most of the local infrastructural activities. In the initial phases, the project was managed by CNR, after which it was passed to INAF. Three institutions belonging to INAF made the technical and managerial efforts: IRA, OAC and OAA. The Nuraghe control system is one of the creations of the INAF project named DISCOS, aimed at providing the Italian radio telescopes with unified managing software.

The contract for the construction of the telescope structure was awarded to MT Mechatronics GmbH, Mainz, Germany.

The authors acknowledge Joseph Schwartz for his valuable suggestions that improved the clarity and readability of the text.

## References

Ambrosini, R., Asmar, S. W., Bolli, P. & Flamini, E. [2011a] *Proc. IEEE* **99**(5), 875–880.

Ambrosini, R. *et al.* [2011b] "The new time and frequency laboratory for the Sardinia Radio Telescope", in *Proc. URSI XXX General Assembly and Scientific Symp.*, A03.9.

Ambrosini, R. *et al.* [2013] *IEEE Int. Conf. Electromagnetics in Advanced Applications*, Torino, Italy, September 9–13.

Bolli, P. & Grueff, G. [2003] *Electron. Lett.* **39**(5), 416–417.

Bolli, P. *et al.* [2013] *IEEE Antennas Propag. Mag.* **55**(5), 19–24.

Bolli, P., Olmi, L., Roda, J. & Zacchiroli, G. [2014] *IEEE Antennas Wireless Propag. Lett.* **13**(1), 1713–1716.

Buttu, M. *et al.* [2012] "Diving into the Sardinia Radio Telescope minor servo system", in *Proc. SPIE Software and Cyberinfrastructure for Astronomy II*, Vol. 8451, 84512L.

Cortès-Medellin, G. [2002] "The 64m Sardinia Radio Telescope Optics Design", in *IEEE Int. Symp. Antennas and Propagation Digest*, pp. 136–139.

Guiar, C. N. *et al.* [1986] "Antenna pointing systematic error model derivations", in *Telecommunications and Data Acquisition Progress Report*, 42-88, pp. 36–46.

Grueff, G. *et al.* [2004] "Sardinia Radio Telescope: The new Italian project", in *Proc. SPIE Ground based Telescopes*, Vol. 5489, p. 773.

Maneri, E. & Gawronski, W. [2002] *IEEE Antennas Propag. Mag.* **44**(4), 40–50.

Melis, A. *et al.* [2014] "An infrastructure for multi back-end observations with the Sardinia Radio Telescope", in *Proc. SPIE. Millimeter, Submillimeter, and Far-Infrared Detectors and Instrumentation for Astronomy VII*, Vol. 9153, 91532M.

Nesti, R. *et al.* [2010] SRT Technical Memo GAI04-TM-13.0.

Orfei, A. *et al.* [2004] "Active surface system for the new Sardinia Radiotelescope", in *Proc. SPIE Astronomical Structures and Mechanisms Technology*, Vol. 5495, 116–125.

Orfei, A. *et al.* [2010] *IEEE Antennas Propag. Mag.* **52**(4), 62.

Orfei, A. *et al.* [2011] SRT Final Report GAI04-FR-5.0 (in Italian).

Orlati, A. *et al.* [2012] "The control software for the Sardinia Radio Telescope", in *Proc. SPIE Software and Cyberinfrastructure for Astronomy II*, Vol. 8451, 84512M.

Ott, M. [1994] **284**, 331–339.

Peverini, O. A., Virone, G., Addamo, G. & Tascone, G. [2011] *IET Microwaves, Antennas Propag.* **5**(8), 1008–1015.

Pisanu, T. *et al.* [2012] "Architecture of the metrology for the SRT", in *Proc. SPIE, Ground-based and Airborne Telescopes IV*, Vol. 8444, 84442E.

Poloni, M. [2010] SRT Technical Memo GAI04-TM-16.0 (in Italian).

Prestage, R. M. *et al.* [2004] "The GBT precision telescope control system", in *Proc. SPIE Ground-based Telescopes*, Vol. 5489, pp. 1029–1040.

Ruze, J. [1952] *Nuovo Cimento Suppl.* **9**(3), 364–380.

Serra, G. *et al.* [2012] "The microwave holography system for the Sardinia Radio Telescope", in *Proc. SPIE Ground-based and Airborne Telescopes IV*, Vol. 8444, 84445W.

Süss, M., Koch, D. & Paluszek, H. [2012] "The Sardinia Radio Telescope (SRT) optical alignment", in *Proc. SPIE Ground-based and Airborne Telescopes IV*, Vol. 8444, 84442G.

Tofani, G. *et al.* [2008] "Status of the Sardinia Radio Telescope project", in *Proc. SPIE Ground-based and Airborne Telescopes II*, Vol. 7012, 70120F.

Valente, G. *et al.* [2010] "The dual-band LP feed system for the Sardinia Radio Telescope prime focus", in *Proc. SPIE Millimeter, Submillimeter, and Far-Infrared Detectors and Instrumentation for Astronomy V*, Vol. 7741, 774126.